\documentstyle[pra,aps,eqsecnum,multicol,epsf]{revtex}

\begin{document}

\title{Comparative study of the critical behavior in one-dimensional\\
random and aperiodic environments\footnote{Dedicated to
  J. Zittartz on the occasion of his 60th birthday}}

\author{Ferenc Igl\'oi$^{1,4}$, Dragi Karevski$^2$ and Heiko
  Rieger$^{3,4}$}

\address{
$^1$ Research Institute for Solid State Physics, 
H-1525 Budapest, P.O.Box 49, Hungary\footnote{Permanent address also:
        Institute for Theoretical Physics,
        Szeged University, H-6720 Szeged, Hungary}\\
$^2$ Laboratoire de Physique des Materiaux, 
Universit\'e Henri Poincar\'e,
F-54506 Vand\oe uvre les Nancy, France\\
$^3$ Institut f\"ur Theoretische Physik, Universit\"at zu K\"oln, 
     50923 K\"oln, Germany\\
$^4$ HLRZ, Forschungszentrum J\"ulich, 52425 J\"ulich, Germany\\
}

\date{January 24, 1998}

\maketitle

\begin{abstract}
  We consider cooperative processes (quantum spin chains and random
  walks) in one-dimensional fluctuating random and aperiodic
  environments characterized by fluctuating exponents $\omega>0$. At the
  critical point the random and aperiodic systems scale essentially
  anisotropically in a similar fashion: length ($L$) and time ($t$)
  scales are related as $L\sim(\ln t)^{1/\omega}$. Also some critical
  exponents, characterizing the singularities of average quantities, are
  found to be universal functions of $\omega$, whereas some others do
  depend on details of the distribution of the disorder. In the
  off-critical region there is an important difference between the two
  types of environments: in aperiodic systems there are no extra
  (Griffiths)-singularities.
\end{abstract}


\newcommand{\bc}{\begin{center}}
\newcommand{\ec}{\end{center}}
\newcommand{\be}{\begin{equation}}
\newcommand{\ee}{\end{equation}}
\newcommand{\beqn}{\begin{eqnarray}}
\newcommand{\eeqn}{\end{eqnarray}}

\begin{multicols}{2}
\narrowtext

\section{Introduction}

Quenched disorder often has a strong effect on the cooperative
properties of stochastic processes and strongly correlated systems,
especially in one space dimension. For example the diffusion process
in a one-dimensional environment (Sinai's walk) \cite{sinai} becomes
extremely slow, the mean-square displacement behaves like
$[X^2(t)]_{\rm av}\sim\ln^4t$, in contrast to the linear
$t$-dependence in the homogeneous case. 

Ultraslow dynamics has recently also been observed in one-dimensional
random quantum spin system \cite{bigpaper,riegerigloi}.  The origin of
the slow relaxation in these strongly correlated systems is again the
presence of quenched disorder, and in particular the vicinity of a
quantum critical point. Generally the presence of quenched disorder
has a more pronounced effect on quantum phase transitions, which occur
at zero temperature and are driven by quantum fluctuations, than on
the so called classical phase transitions, which are driven by thermal
fluctuations.

Up to now, these two observations, namely the ultraslow diffusion in
general one-dimensional disordered environments and the so called
``quantum''activated dynamics in random quantum spin chains, seemed to
be unrelated. One aim of the present paper is to demonstrate a
intimate connection between both. Another issue is the question in how
far these effects are also present in {\it aperiodic}, i.e.\ 
non-random but nevertheless inhomogeneous environments. As we will
see, relevant aperiodic systems bear a lot of similarities with
completely random systems. However, also crucial differences exist, in
particular in the so called off-critical regime, i.e.\ the
Griffiths-McCoy \cite{griffiths,mccoy} region for the quantum spin
chains and the anomalous diffusion regime of the random walk. It is
simply not existent in aperiodic systems. A short account of the
latter issue (random versus relevant aperiodicity) has been given
elsewhere \cite{aperiodicshort}.

Although the results we report are of rather general validity we have,
for concreteness and for the lack of space, to confine ourselves to
particular examples. So the prototype of a random quantum spin system
is the random transverse Ising model (random TIM), which has received
considerable interest recently
\cite{fisher,mckenzie,youngrieger,profiles,riegerigloi,young,bigpaper,qsg,senthil,fm2d}.
In one dimension the model is defined by the Hamiltonian
\be
H=-\sum_l J_l \sigma_l^x \sigma_{l+1}^x-\sum_l h_l \sigma_l^z\;.
\label{hamilton}
\ee
in terms of the $\sigma_l^x$, $\sigma_l^z$ Pauli matrices at site $l$.
Here the exchange couplings $J_l$ and the transverse fields $h_l$ are
quenched random variables, taken independently from distributions
$\pi(J)$ and $\rho(h)$, respectively. The quantum model in
(\ref{hamilton}) is closely related to a two-dimensional classical
Ising model with randomly layered couplings, which was first
introduced and partially solved by McCoy and Wu
\cite{surfising,mccoywu} (see also
\cite{zittartz,shankar,nieuwenhuizen}). The quantum control parameter of
the model is defined by
\be
\delta={[\ln h]_{\rm av}-[\ln J]_{\rm av} 
\over \rm{var}[\ln h]+\rm{var}[\ln J]}\;.
\label{delta}
\ee
For $\delta<0$ ($\delta>0$) the system is in the ordered (disordered)
phase and at $\delta=0$ there is a phase transition in the system. One
surprising observation is that at the critical point some physical
quantities are not {\it self-averaging}, thus the {\it typical} and
the {\it average} values of those are different. For these observables
the critical properties are determined by the so called {\it rare events},
(which occur with vanishing probability) dominating the mean values.

The average magnetization at the surface, $m_s$ and in the bulk,
$m_b$, of the system vanishes as a power law close to the critical
point as $m_s\sim\delta^{\beta_s}$ and $m_b\sim\delta^\beta$,
respectively, with exponents
\be
\beta_s=1\quad{\rm and}\quad\beta=\frac{3-\sqrt{5}}{2}\;.
\label{beta}
\ee
The average spin-spin autocorrelation function
$G(l)=[\langle\sigma_i^x\sigma_{i+r}^x\rangle]_{\rm av}$ involves the average
correlation length $\xi$, which diverges at the critical point as
$\xi\propto\vert\delta\vert^{-\nu}$ with the correlation length exponent
\be
\nu=2\;,
\label{nurand}
\ee
which differs from the exponent of the {\it typical} correlation
function 
$G_{\rm typ}(l)=\exp([\ln\langle\sigma_i^x\sigma_{i+r}^x\rangle]_{\rm av}\}$, 
which is
\be
\nu_{\rm typ}=1\;.
\label{nutyp}
\ee
The time-dependent correlations of the model are very special, they
differ completely from those in the pure system. The autocorrelation
function 
\be
G_l(t)=[\langle\sigma_l^x(0)\sigma_l^x(t)\rangle]_{\rm av}
\label{autocorr}
\ee
at the critical point decays on a logarithmic scale
\be
G_l(t)\sim(\ln t)^{-\eta}\quad(\delta=0)\;,
\ee
where the decay exponent $\eta$ satisfies the scaling relation
$\eta=\beta/\nu$ and $\eta_s=\beta_s/\nu$ for site $l$ in the bulk and
on the surface, respectively. Leaving the critical point in any
direction one enters the Griffiths-McCoy regions, in which the
connected autocorrelation function has a power law decay:
\be
G_l(t)\sim t^{-z(\delta)}\quad(\delta\ne0)\;,
\ee
where the dynamical exponent $z(\delta)$ is a continuous function of
the parameter $\delta$. Close to the critical point it is
$z(\delta)\approx1/2\delta$\cite{fisher}. As a consequence of the power-law decay
of the autocorrelations in the Griffiths -McCoy phase the magnetization
is a singular function of the uniform magnetic field $H_x$ as $m_{\rm
sing}\propto\vert H_x\vert^{1/z(\delta)}$.

The above results about the random TIM are independent of the actual form of
the probability distribution. It is often argued that the perturbation
caused by the disorder, with respect to the pure system, is connected
to the fluctuating energy per spin:
\be
\Delta(L)/L=\frac{1}{L}
\sum_{l=1}^L\left( J_l-[J]_{\rm av} \right) \sim L^{\omega-1}\;.
\label{fluctuate}
\ee
Here $L$ is the linear size of the sequence of couplings $J_l$ under
consideration and $\omega=\omega_{\rm rand}=1/2$ is the fluctuating or
wandering exponent. The Harris criterion\cite{harris} makes a statement on behalf
of the relevance/irrelevance of the perturbation of the critical
behavior of the pure system by the above disorder. One compares the
strength of the thermal fluctuations at the critical point with the
fluctuating energy in (\ref{fluctuate}) on a length scale $L$
identical to the correlation length. This yields in a one-dimensional
system $\Phi=1+\nu(\omega-1)$ for the cross-over exponent\cite{luck,igloi93}
and indeed,
for the random sequence the perturbation is relevant, since with $\nu=1$ and
$\omega=1/2$ one gets $\Phi=1/2>0$.

It is known that there are non-random deterministic sequences,
generated through substitutional rules, which have unbounded
fluctuations, so that the corresponding fluctuating exponent is
$\omega>0$. Having this similarity between random and aperiodic
sequences in mind, one might ask the question whether or not the
fluctuating exponent $\omega$ is the only quantity that determines the
critical behavior of systems with unbounded fluctuations in the
couplings.

As an example we consider the Rudin--Shapiro (RS) sequence\cite{dekking},
which is
built on four letters {\bf A},{\bf B},{\bf C} and {\bf D} with the
substitutional rule:
\be
{\bf A} \to {\bf AB}~~,~~{\bf B} \to {\bf AC}~~,~~{\bf C} \to {\bf DB}~~,~~
{\bf D} \to {\bf DC}\;.
\label{rs}
\ee
Thus starting with a letter {\bf A} one proceeds as: ${\bf A} \to {\bf
AB} \to {\bf ABAC} \to {\bf ABACABDB} \to$ etc., and one may assign
different couplings to the different letters. The fluctuating exponent
of the RS-sequence, which was originally introduced to mimic random
fluctuations, $\omega_{\rm RS}=1/2$, i.e.\ just the same as for the
random sequence.

We note that up to now the critical behavior of aperiodic systems has
been studied mainly for such distributions that are {\it non-relevant}
in the sense of the Harris criterion (\i.e.\ $\omega\le 0$). Especially
for marginal sequences (i.e.\ $\omega=0$) non-universal critical
behavior coupling dependent anisotropy exponents have been found in
exact calculations for the transverse field Ising
chain\cite{bercheberche,igloiturban96,itks,grimmbaake}.

In the relevant case, $\omega>0$, there are only a few exact results,
obtained for one specific representation described after eq(\ref{rs})
starting with the letter {\bf A}. For this case the magnetization of
finite RS-chains of length $L$ at the critical point has been shown to
behave at the two end points as\cite{igloiturban94}
\be
m_s(RS)\sim\exp(-{\rm const}\cdot\sqrt{L})
\label{msrs}
\ee
and $\overline{m_s}(RS)={\rm const}>0$. Moreover, the lowest
excitation energy at the critical point is given by
\be
\varepsilon_1(L)\sim\exp(-{\rm const}\cdot\sqrt{L})\;.
\label{eps}
\ee
We also note on recent Monte-Carlo simulations on the critical behavior
of the $Q=8$-state Potts model with layered aperiodic modulations\cite{bcb}.

In the present paper we are going to study the {\it average
quantities} also for aperiodic chains. This means that for instance
for the bulk magnetization we average over all sites of the
system. This averaging process, however, could be performed in another
way. One generates (hypothetically) an infinite sequence through
substitution, cut out segments of length $L$ starting at all different
positions of the (infinite) sequence and then average over all
segments. Then, considering the example of surface magnetization one
could expect that the two generic situations in and below eq(\ref{msrs})
will appear with
given probabilities, which are then connected to the average critical
behavior of the system.

The structure of the paper is the following: in sec.\ II we present
the basic notations of aperiodic sequences used in the paper. The free
fermionic description of the TIM is given in sec.\ III. Results about
the TIM both at the critical point and in the off-critical region are
presented in sec. IV and V, respectively. We conclude with a final section on a parallel
analysis of the behavior of a random walk in random or aperiodic
environments.

\section{Aperiodic substitutional sequences}

In this paper we consider sequences generated via substitution on a
finite alphabet such that, in the case of two letters {\bf A} and {\bf
B}, one substitutes ${\bf A}\to S({\bf A})$ and ${\bf B}\to S({\bf
B})$. For the RS-sequence the explicit relations are given in
eq(\ref{rs}). The properties of the sequence are governed by the
substitutional matrix:
\be
{\bf M}=\left(
\begin{array}{cc}
n_{\bf A}^{S({\bf A})} & n_{\bf A}^{S({\bf B})} \\
n_{\bf B}^{S({\bf A})} & n_{\bf B}^{S({\bf B})} \\
\end{array}\;
\right) 
\ee 
where the matrix element $n_{\bf i}^{S({\bf j})}$ is the number of the
letters ${\bf i}$ occurring in the substitution $S({\bf j})$. The
matrix elements of ${\bf M}^n$ contain the same numbers in the
sequence after $n$ iterations.

If ${\bf U}_r$ denotes the right eigenvector of ${\bf M}$ with eigenvalues
$\Omega_r$, the asymptotic density of ${\bf i}$ is given by
\be
\rho_\infty^{({\bf i})}=
\frac{{\bf U}_1({\bf i})}{\sum_{\bf j}{\bf U}_1({\bf j})}\;,
\ee
where ${\bf U}_1$ is the eigenvector corresponding to the leading
eigenvector $\Omega_1$. The length of the sequence after $n$ iterations
is related to the leading eigenvalue through 
\be
L_n\sim\Omega_1^n\;.
\label{length}
\ee

We consider the TIM in (\ref{hamilton}) with a constant transverse
field $h_l=h$ in an aperiodic environment. We assign to each letter in
the sequence a coupling constant (for example $J_{\bf A}$ and $J_{\bf
B}$). The cumulated deviation $\Delta(L)$ from the average coupling
$[J]_{\rm av}$ in eq(\ref{fluctuate}) scales with $L$ as
\be
\Delta(L)\propto\vert\Omega_2\vert^n\propto L^\omega\;,
\label{fluctuate2}
\ee
where $\Omega_2$ is the next-to-leading eigenvalue of the substitution
matrix ${\rm M}$ and $\omega$ is the fluctuating exponent already
defined in (\ref{fluctuate}). Combining (\ref{fluctuate2})
and(\ref{length}) one gets
\be
\omega=\frac{\ln\vert\Omega_2\vert}{\ln\Omega_1}\;.
\label{fluctexp}
\ee
The critical point $h_c$ of the system is obtained from the relation
(\ref{delta}) with  $\delta=0$ as
\be
\sum_{\bf i}\rho_\infty^{({\bf i})}\ln J_{\bf i}=\ln h_c
\label{critpoint}
\ee

In the following we introduce a family of relevant sequences defined
on $k$-letters ($k$-general sequence) ${\bf A}_1, {\bf A}_2, \ldots,
{\bf A}_k$ with the substitutional rules:
\be
\begin{array}{cclcrcl}
{\bf A}_1 & \to & {\bf A}_1 {\bf A}_2         &                     \\
{\bf A}_i & \to & {\bf A}_{i-1} {\bf A}_{i+1} & \qquad & 1&<i\le&k/2 \\
{\bf A}_i & \to & {\bf A}_{i+1} {\bf A}_{i-1} & \qquad & k/2&<i<&k    \\
{\bf A}_k & \to & {\bf A}_k {\bf A}_{k-1}     &        &    &   &             
\end{array}
\label{kseq}
\ee
The substitutional matrix of the sequence is given by
\be
{\bf M}=
\left(\begin{array}{ccccccc}
1 & 1 &        &        &        &   &   \\
1 & 0 & 1      &        &        &   &   \\
  & 1 & 0      & 1      &        &   &   \\ 
  &   & \ddots & \ddots & \ddots &   &   \\ 
  &   &        &        &    1   & 0 & 1 \\
  &   &        &        &        & 1 & 1 
\end{array}
\right)
\ee
with the eigenvalues
\be
\Omega_r=2\cos[\frac{\pi}{k}(r-1)]\quad r=1,2,\ldots,k\;.
\ee
Thus the fluctuating exponent $\omega$ from (\ref{fluctexp}) for
$k$-general sequences is given by
\be
\omega_k=\frac{\ln[2\cos(\pi/k)]}{\ln 2}
\label{omegak}
\ee
and is positive for $k\ge4$. The leading right eigenvector of ${\bf M}$
is just ${\bf U}_1({\bf i})={\rm const}$, thus the asymptotic density
of letters is given by
\be
\rho_\infty^{({\bf i})}=1/k\;.
\label{density}
\ee
We note that the $k$-general sequence for $k=2$ is just the Thue-Morse
sequence, while for $k=4$ we recover the RS-sequence mentioned in the
introduction in (\ref{rs}).

To complete this section on the introduction of aperiodic sequences we
have to discuss the averaging process. As described in the
introduction the average for aperiodic sequences is made in such a way
that from a (hypothetical) infinite sequence segments of length $L$
are cut out starting at all different points in the infinite sequence,
these we call the realizations of length $L$ of the aperiodic
sequence.  Then we average physical observables over these
realizations. The number of different realizations $R(L)$ of a {\it
random} (binary) sequence grows exponentially with the size of the
system, whereas in the {\it aperiodic} case have a much slower, namely
linear increase:
\be
R(L)=aL
\ee
with $a$ being some constant. For the RS-chain the number of different
realizations is less than $16L$, and one can obtain the exact average
value by the following procedure. Consider the RS-sequence in
(\ref{rs}) generated from a letter {\bf A}, take the first $4L$
sequels of length $L$, starting at positions $1,2,\dots,4L$ and take
also their reflection symmetric counterparts.  Then repeat the
procedure starting with a letter {\bf D}.  The averaging then should
be performed over these $16L$ realizations. Therefore it is possible in
numerical computations to do an {\it exact} average for the
RS-sequence, taking into account {\it all} different realizations and
to obtain numerically exact results for relatively large ($L \le 512$)
aperiodic chains.  For other values of $k$ in the $k$-general case the
number of different realizations is still linear in $L$, however, no
similar simple rule to generate all of them in linear computing time
is known to us.

\section{Free fermionic representation of the TIM}

We consider the TIM in eq(\ref{hamilton}) on a finite chain of length
$L$ with free or fixed boundary conditions, thus we set $J_L=0$.  The
simplest way to calculate various physical quantities of the model is
to transform it into a fermionic representation according to Lieb,
Schultz and Mattis \cite{fermion}. A detailed description of the method
and its application to different physical quantities can be found in
\cite{bigpaper}, hereafter referred to as paper I. Here we briefly
recapitulate the main results.

The Hamiltonian (\ref{hamilton}) is expressed in terms of fermion
creation and annihilation operators $\eta_q^+$ and $\eta_q$,
respectively:
\be
H=\sum_{q=1}^L\varepsilon_q\left(\eta_q^+\eta_q-\frac{1}{2}\right)\;,
\label{fermion}
\ee
where the fermionic energies $\varepsilon_q$ are identical to the
non-negative eigenvalues of the $2L\times2L$ tridiagonal matrix 
${\bf T}$ with eigenvectors $V_q$:
\end{multicols}
\widetext
\noindent\rule{20.5pc}{.1mm}\rule{.1mm}{2mm}\hfill
\be
{\bf T} = \left(
\matrix{
 0  & h_1 &     &       &       &       &     \cr
h_1 &  0  & J_1 &       &       &       &     \cr
    & J_1 &  0  & h_2   &       &       &     \cr
    &     & h_2 &  0    &\ddots &       &     \cr
    &     &      &\ddots&\ddots &J_{L-1}&     \cr
    &     &      &      &J_{L-1}&   0   & h_L \cr
    &     &      &      &       &  h_L  &  0  \cr}
\right)\quad,\qquad
V_q = \left(\matrix{
-&\Phi_q(1)\cr
 &\Psi_q(1)\cr
-&\Phi_q(2)\cr
 &\vdots\cr
 &\Psi_q(L-1)\cr
-&\Phi_q(L)\cr
 &\Psi_q(L)}
\right)\quad.
\label{trid}
\ee
\hfill\rule[-2mm]{.1mm}{2mm}\rule{20.5pc}{.1mm}
\begin{multicols}{2} 
\narrowtext
\noindent 
The magnetization profiles
\be
m_l=\langle0\vert\sigma_l^x\vert0\rangle
\label{magl}
\ee
for symmetry braking boundary conditions (b.c.) can be expressed as
$l\times l$ determinants, whose elements are defined in terms of the
components $\Psi_q(i)$ and $\Phi_q(i)$ of the eigenvectors $V_q$ in
(\ref{trid}). If we fix the spin at one end of the chain, which is
equivalent to setting $h_L=0$, the local magnetization is given by
\be
m_l^{\rm free}=\left|\,\matrix{
H_1&G_{11}&G_{12}&\ldots&G_{1l-1}\cr
H_2&G_{21}&G_{22}&\ldots&G_{2l-1}\cr
\vdots&\vdots&\vdots&\ddots&\vdots\cr
H_l&G_{l1}&G_{l2}&\ldots&G_{ll-1}\cr}
\right|\;,
\label{determinant}
\ee
where
\beqn
H_j   &=&\Phi_1(j)\nonumber\\
G_{jk}&=&-\sum_{q} \Psi_q(k) \Phi_q(j)\;.
\label{matrelm}
\eeqn
The surface magnetization for fixed-spin b.c.\ $m_s=m_1$ is given by
the very simple formula
\be
m_s=\left[1+\sum_{l=1}^{L-1} \prod_{j=1}^l
\left( h_j \over J_j \right)^2\right]^{-1/2}\;,
\label{surfmag}
\ee
which is {\it exact} for any {\it finite} system. 

The magnetization profile can be calculated analogously (see paper I),
when both surface spins are fixed (++ and +-- b.c.s). Finally , for a
finite open chain the ground state expectation value of $\sigma_l^x$
in (\ref{magl}) is identically zero, due to symmetry. Here, as shown
in \cite{profiles} the matrix-element profile
\be
m_l^{\rm free}=\langle1\vert\sigma_l^x\vert0\rangle
\label{magfree}
\ee
is of importance, which is formally given by eq(\ref{determinant})
with $h_L\ne0$.

Here we note on a simple estimate for the excitation energy $\varepsilon_1(L)$
in a open chain of length $L$\cite{itks}:
\be 
\varepsilon_1(L) \sim m_s \overline{m} _s
\,\,h_L\prod_{i=1}^{L-1} {h_i\over J_i}\;.
\label{epsform}
\ee
provided $\lim_{L \to \infty} \varepsilon_1(L) L=0$. Here $m_s$ and
$\overline{m}_s$ denote the finite size surface magnetizations at both
ends of the chain, as defined in eq(\ref{surfmag}) (for $\overline{m}_s$
simply replacing $h_j/J_j$ by $h_{L-j}/J_{L-j}$ in this equation).

Next we consider the dynamical correlation functions of the system as a
function of the imaginary time $\tau$:
\be
G_l(\tau)=\langle0\vert\sigma_l^x(\tau)\sigma_l^x(0)\vert0\rangle\;.
\label{corr}
\ee
For surface spins this can be expressed in the simple form
\be
G_1(\tau)=\sum_q | \Phi_q(1)|^2 \exp(-\tau \varepsilon_q)\;.
\label{surfcorr}
\ee
whereas in the bulk $G_l(\tau)$ can be expressed as a Pfaffian that can
be evaluated via a determinant of $l\times l$ antisymmetric matrix
(see paper I for details).

\section{Critical Properties}

In what follows we take for the Hamiltonian (\ref{hamilton}) homogeneous
fields $h_l=h$ and two-valued couplings, say $J_l=\lambda$ and
$J_l=1/\lambda$. For the $k$-general sequence in (\ref{kseq})
$J_l=\lambda$ for letters $\bf{A_i}$ with $i<(k+1)/2$ and $J_l=1/\lambda$
for letters with $i>(k+1)/2$. For an odd $k$ we take $J_l=1$ for $i=(k+1)/2$.
Thus for the RS sequence in (\ref{rs}) we have $J_{A}=J_{B}=\lambda$ and
$J_C=J_D=1/\lambda$. Then from (\ref{critpoint}) and (\ref{density}) for
the critical point follows:
\be
\delta=\ln h_c=0\;,
\label{apercrit}
\ee
independently of $k$. Similarly, for the random model we take homogeneous
fields and binary distribution of the couplings: $J_l=\lambda$ and
$J_l=1/\lambda$, with the same probability. The critical point in this
case is also given by eq(\ref{apercrit}).

\subsection{The distribution of low energy excitations}

The basic features of the random transverse Ising chain at criticality
have been described in the introduction. Here we stress again that
many unusual properties of the random TIM are connected to the extremely
broad distribution of the different physical quantities at the
critical point. Concerning the energy gap $\Delta E$ in a finite
system of size $L$ the appropriate scaling variable is $\ln\Delta
E/\sqrt{L}$. The reason for this is that the system is essentially
anisotropic at criticality, which means that times and length scale
are related in an exponential rather than algebraic manner, c.f.\
eq(\ref{eps}).

For relevant aperiodic systems (i.e.\ those with a positive
fluctuating exponent $\omega$) one expects from scaling considerations
\cite{luck,grimmbaake,igloiturban94,igloi93} that the energy gap at the
critical point has a similar scaling relation as in random systems:
\be
\Delta E(L) \sim \exp(-{\rm const}\cdot L^{\omega})\;.
\label{apdelta}
\ee
This form actually follows from the formula in eq(\ref{epsform})
for the energy gap of a sample with local order at the two ends, i.e.\
$m_s={\cal O}(1)$ and $\overline{m}_s={\cal O}(1)$, where $m_s$ and
$\overline{m}_s$ are the surface magnetizations at the left and right
end, respectively, to be calculated with eq(\ref{surfmag}). Then we
have $\varepsilon_1\sim\prod_{i=1}^{l-1} {h_i\over
J_i}\sim\exp(-l_{\rm tr} \overline{\ln(J/h)})$, where $l_{\rm tr}\sim
L^\omega$ measures the size of the transverse fluctuations and
$\overline{\ln(J/h)}$ is an averaged reduced coupling, from which
(\ref{apdelta}) follows.

To check the validity of the above scaling relation (\ref{apdelta}) we
have investigated the probability distribution of the energy gap
$P_L(\Delta E)$ at the critical point of the RS-chain and compared it
with the same quantity for the random chain. 

\begin{figure}
\epsfxsize=\columnwidth\epsfbox{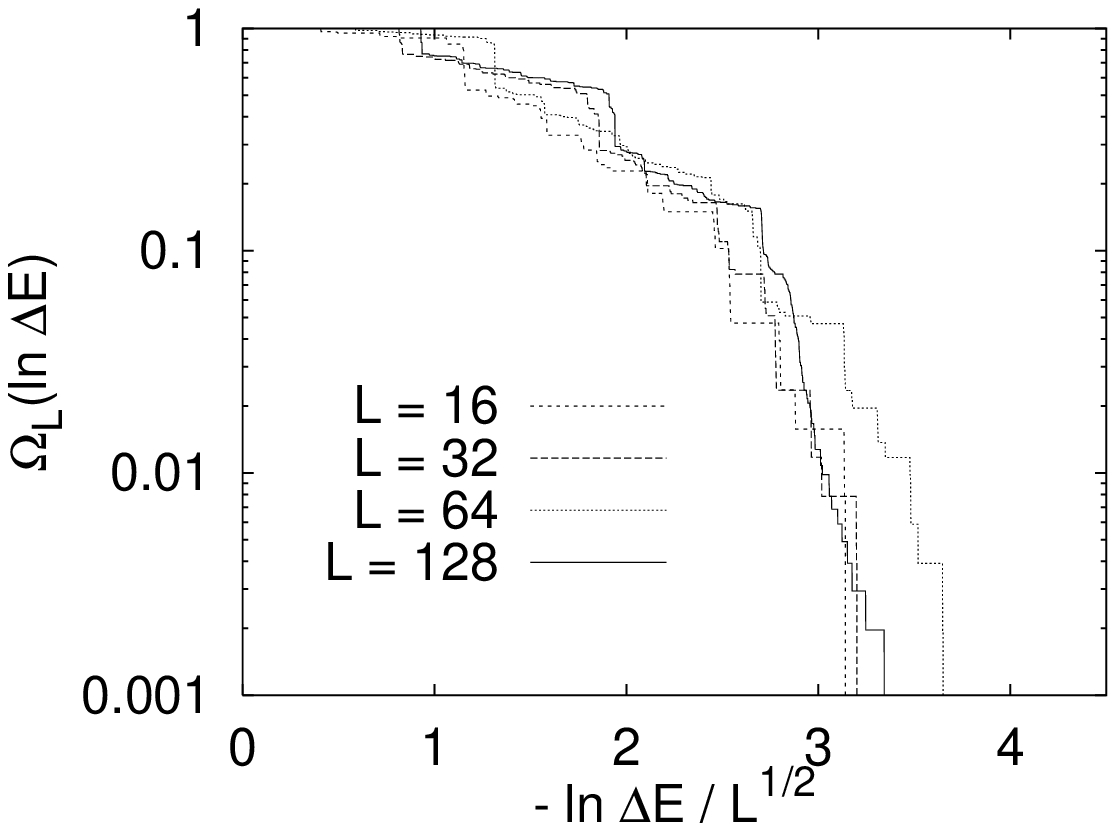}
\epsfxsize=\columnwidth\epsfbox{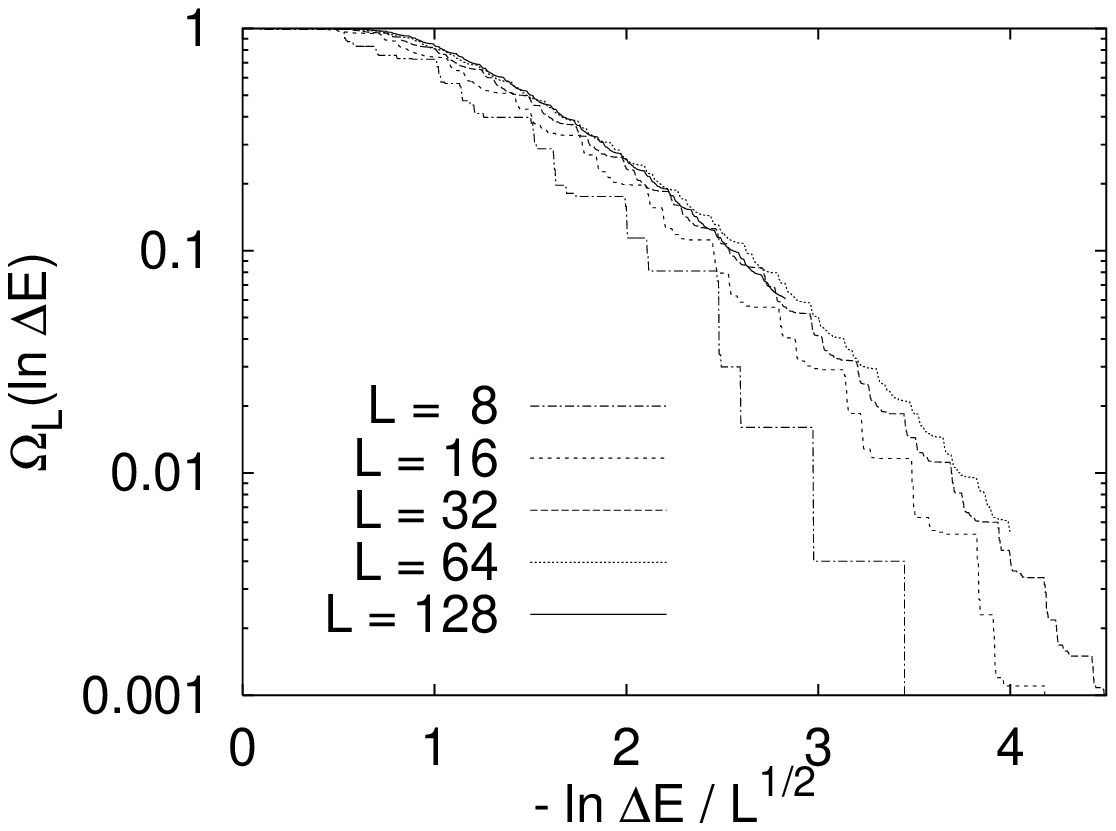}
\caption{{\bf a)} (Top) Scaling plot of the integrated probability density
  $\Omega(\ln\Delta E)$ versus the scaling variable $(\ln\Delta
  E/L^{1/2})$ for the RS sequence (exact average) with $\lambda=4$,
  c.f.\ eq.(\protect{\ref{accumulated}}).  {\bf b)} (Bottom) The same
  as in a) for the the random sequence.}
\label{fig1}
\end{figure}

According to eq(\ref{apdelta}) the appropriate scaling variable is
$\ln\Delta E/L^{1/2}$ for both systems. Indeed, as seen in Fig.\ 1a,
the integrate probability distribution function
\be
\Omega_L(\ln \Delta E)=\int_{-\infty}^{\ln\Delta E} \!\!dy\,
\tilde{P}_L(y)\sim\tilde{\Omega}(\ln \Delta E /\sqrt{L})
\label{accumulated}
\ee
with $\tilde{P}_L(\ln\Delta E)=P_L(\Delta E)\Delta E$, has a good data
collapse using this reduced variable. Considering the same quantity
for the random chain one can observe the same scaling behavior, as
shown in Fig.1b. Thus we can conclude, that both the random and the
RS-chains have logarithmically broad distribution of the energy gaps
at the critical point, from this fact one expects similar consequences
for the critical behavior in the two systems.


\subsection{Surface magnetization}

For the TIM the surface magnetization represents perhaps the most
simple order-parameter of the system, which in the fix-free b.c.\ is
explicitly given by the simple formula in (\ref{surfmag}). For one
single RS-chain, starting with the letter {\bf A} this expression has
been exactly evaluated at the critical point with the following
results\cite{igloiturban94}.

For $\lambda>1$, when the average couplings at the surface are
stronger than in the bulk, the surface is ordered at the critical
point, so that the surface magnetization is finite:
\be
\lim_{L\to\infty}m_s(L,\lambda,h=1)=
\frac{\lambda^2-1}{\lambda^4-\lambda^2+1}\qquad\lambda>1
\ee
and approaches unity in the limit $\lambda\to\infty$. On the other
hand for $\lambda<1$, when the average couplings at the 
surface are weaker than in the bulk the critical
surface magnetization vanishes as 
\be
m_s(L,\lambda,h=1)\sim\exp(-{\rm const}\cdot\sqrt{L})\;.
\label{surfmagexp}
\ee
Proceeding by studying the critical point magnetizations in other
sequels of the RS-chain one can notice that the above two examples are
generic: in a sample the surface magnetization is either finite
$m_s={\cal O}(1)$, or it vanishes in the stretched exponential form as
in eq(\ref{surfmagexp}). Therefore the average is dominated by the
sample with finite surface magnetization (``rare events''), which
occur with probability $P_{\rm rare}(L)\propto L^{-\gamma}$. Thus the
critical surface magnetization is {\it not self-averaging}, it is
determined by rare events and its scaling dimension $x_m^s$ defined by
the asymptotic relation $[m_s(L,h)]_{\rm av}\sim L^{-x_m}$ is just
$x_m^s=\gamma$.

In the following we make extended use of this observation and
calculated $x_m^s$ exactly. Here we adopt the random walk picture of
paper I. {\it First} we assign to each sample with a given realization
of couplings a walk, which starts at the origin and makes the $l$-th
step $+1$ ($-1$) for a coupling $J_l=\lambda$ ($J_l=1/\lambda$).
{\it~Second}, we take the limit $\lambda\to\infty$, in which only
those samples have non-vanishing surface magnetization, where the
corresponding walk never visits sites with negative coordinates. Thus
the proportion of rare events is given by the survival probability of
the walk $P_{\rm rare}(L)=P_{\rm surv}(L)$. Thus the {\it third} point
in the study is to calculate the surviving probability of the walker.
For the random chain the corresponding random walk is characterized by
a surviving probability of $P_{\rm surv}\sim L^{-1/2}$, thus one gets
the exact result:
\be
x_m^s({\rm rand})=1/2\;.
\label{xmsrand}
\ee

For the RS-chain we have performed the exact analysis of the average
surface magnetization, the result of which is presented in the
Appendix. The leading $L$-dependence of it is given by
\beqn
[m_s(L,\lambda\to\infty,h=1)]_{\rm av}
&=&\frac{5}{8}\left(\frac{1}{\sqrt{2}}+\frac{1}{4}\right)\cdot L^{-1/2}
\nonumber \\
&+&{\cal O}(L^{-3/4})
\label{rslimit}
\eeqn
from which the value of the scaling dimension 
\be
x_m^s({\rm RS})=1/2\;.
\label{xmsrs}
\ee
follows. Thus we conclude that the surface magnetization scaling
dimension is the same for the random and the RS-chain.

In what follows we show that for relevant aperiodic sequences $x_m^s$
is a simple function of the fluctuating exponent $\omega$. We consider
the scaling behavior of the surviving probability $P_{\rm surv}(L)$ of
the corresponding aperiodic walk, performing a discrete scale
transformation, which corresponds to a substitutional step of the
sequence. Then the length of the walk scales as $L\to L\Omega_1$,
whereas the transverse fluctuations $l_{\rm tr}$ scale like $l_{\rm
tr}\to l_{\rm tr}\vert\Omega_2\vert$. Then the number of surviving
walks with $L$ steps, $N(L)$, scales as $N(L)\to N(L\Omega_1)=
\vert\Omega_2\vert N(L)$, since the number of these walks is
proportional to the size of the transverse fluctuations. Remembering
that the total number of different sequels is $R(L)=aL$ the survival
probability $P_{\rm surv}(L)=N(L)/R(L)$ satisfies the scaling relation
\be
P_{\rm surv}(\Omega_1 L)=\frac{\vert\Omega_2\vert}{\Omega_1}
P_{\rm surv}(L)\quad\Rightarrow P_{\rm surv}(L)\propto L^{-(1-\omega)}
\ee
from which the value of the surface magnetization scaling dimension
can be read off as
\be
x_m^s=1-\omega\;.
\label{xmsall}
\ee
We have seen by an exact analytical treatment that this relation is
satisfied for the random and the RS-chain. For the first few members
of the family of $k$-general sequences we verified eq(\ref{xmsall})
numerically and obtained $x_m^s(k=5)=0.3059(5)$ and $x_m^s(k=6)=0.2076(4)$,
which is in good agreement with the corresponding prediction from
(\ref{xmsall}) $x_m^s(k=5)=0.3058$ and $x_m^s(k=6)=0.2075$,
respectively.

Now we follow the analysis of paper I and calculate the correlation
length critical exponent $\nu$ from the $\delta=\ln h$ dependence of the
surface magnetization. In the scaling limit $L \gg 1$, $|\delta| \ll
1$ the surface magnetization can be written as $[m_s(L,\delta)]_{\rm
av}=[m_s(L,0)]_{\rm av} \tilde{m_s}(\delta L^{1/\nu})$.  Expanding the
scaling function into a Taylor series $\tilde{m_s}(z)=1+bz+O(z^2)$ one
obtains for the $\delta$ correction to the surface magnetization:
\be
[m_s(L,\delta)]_{\rm av}-[m_s(L,0)]_{\rm av} \sim \delta L^{\Theta}\;.
\label{tempmag1}
\ee
with 
\be
\Theta=1/\nu-x_m^s\;. 
\label{theta}
\ee
This exponent can also be determined exactly in the $\lambda \to
\infty$ limit from random walk arguments. As shown in
paper I the surface magnetization of {\it rare events} is
given by:
\be
m_s(L,\delta)=(1+n)^{-1/2}-\delta {\sum_{i=1}^n l_i \over (n+1)^{3/2}}
+O(\delta^2)\;,
\label{tempcorr}
\ee
where the corresponding surviving walk returns $n$-times to its
starting point after $l_1,l_2,\dots,l_n$ steps. The largest
contribution to the coefficient of the term proportional to $\delta$
in (\ref{tempcorr}) comes from those surviving walks, that have a
large $n$, i.e.\ which visit the starting point frequently. For
general aperiodic sequences $n$ grows with the length of the walk as
\be
n\sim L^{d_s}
\label{return}
\ee
and we call $d_s$ the surface fractal dimension of the surviving
walks. One can check for the $k$-general sequences that after two
substitutional steps, when $L\to4L$, the number of return points
scales as $n\to2n$, thus from eq(\ref{return}) we get
\be
d_s=1/2
\ee
independent of the value $k\ge4$.

Next we are going to perform the average of the linear term in
(\ref{tempcorr}). Here we note that among the $R(L)$ different samples
there are $D(L)=O(n)$ that deliver the same dominant contribution: each
of those has $N_R(L)=O(n)$ return points of characteristic length
$l_i\sim l_{\rm char}(L)={\cal O}(L)$. Thus the average of the linear
term in (\ref{tempcorr}) grows like $D(L)\cdot
l_{\rm char}(L)\cdot N_R(L)/(R(L)\cdot n^{3/2})\sim L^{d_s/2}$. Hence,
comparing with
(\ref{tempmag1}) one gets the exponent relation:
\be {1 \over\nu}-x_m^s={d_s \over 2}\;.
\label{exprel}
\ee
The random case $\nu({\rm rand})=2$ is formally contained in
(\ref{exprel}) with $d_s=0$, since a surviving random walk returns
$n=O(1)$-times to the starting point. For the family of $k$-general
sequences $d_s=1/2$ for all values of $k$, thus the corresponding
correlation length exponent is $\nu(k)=4/(5-4\omega_k)$, with
$\omega_k$ given in (\ref{omegak}).  In particular for the
RS-sequence we get
\be
\nu(RS)=4/3\;,
\label{nurs}
\ee
We have checked this relation numerically by evaluating exactly the
average surface magnetization up to $L=2^{21}$. The two point fits for
the exponent $\Theta$ in (\ref{tempmag1}) comparing systems of size
$2^{2l-1}$ and $2^{2l+1}$ are given in table I. Using standard
sequence extrapolation techniques\cite{BST} on the data in Table I we got an
estimate $\Theta=0.2501(2)$ which is in excellent agreement with the
scaling result (\ref{theta}) with (\ref{nurs}).
\bc
\begin{tabular}{|c|c|}
\hline
$l$ & $\Theta$\\
\hline
2 & .19172235\\
3 & .15877113\\
4 & .16048277\\
5 & .17656409\\
6 & .19500760\\
7 & .21068687\\
8 & .22247716\\
9 & .23086745\\
10& .23670532\\
\hline
\end{tabular}
\ec
TABLE I: Numerical estimates of the exponent $\Theta$ in
(\ref{tempmag1}) comparing numerically exact results on finite systems
of sizes $2^{2l-1}$ and $2^{2l+1}$.
\bigskip

For the other members of the $k$-general sequence the numerical estimates
are $\nu(k=5)=1.82(5)$ and $\nu(k=6)=2.23(5)$, what should be compared with
the scaling predictions $\nu(k=5)=1.799$ and $\nu(k=6)=2.186$, obtained from
eq(\ref{exprel}).

\subsection{Magnetization profiles}

In a geometrically constrained finite system at the critical point the
appropriate way to describe the position dependent physical
quantities, such as magnetization or energy density, to use density
profiles rather then bulk and surface observables\cite{fisherdegennes}.
For two-dimensional
classical and one-dimensional quantum systems with homogeneous
couplings conformal invariance provides a useful tool to study various
geometries. Let us consider a critical system confined between two
parallel plates, which are at large, but finite distance $L$ apart,
where the local densities $\phi(r)$ vary with the distance $l$ from
the plates as a smooth function of $l/L$. According to conformal
invariance\cite{burkhardt}
\be
\langle\phi(l)\rangle_{ab}=\left[
\frac{L}{\pi}\sin\pi\frac{l}{L}\right]^{-x_\phi}
G_{ab}(l/L)\;,
\label{diag}
\ee
where $x_\phi$ is the scaling dimension of the operator $\phi$ and the
scaling function $G_{ab}(x)$ depends on the universality class of the
model and on the type of the boundary conditions to the left $a$ and
to the right $b$. With symmetric b.c.\ the scaling function is
constant $G_{aa}=A$.

In two dimensions conformal invariance can also be used to predict the
critical off-diagonal matrix elements profiles
$\langle\phi\vert\phi(l)\vert0\rangle$, where $\langle\phi\vert$
denotes the lowest excited state leading to a non-vanishing matrix
element (see \cite{profiles,bigpaper}). These off-diagonal profiles
give information about the surface and bulk critical behavior via
finite size scaling.  With symmetric b.c.\ one obtains for the 
profile\cite{turbanigloi97}
\be
\langle \phi\vert\phi(l)\vert0\rangle
\left(\frac{\pi}{\L}\right)^{x_\phi}
\left[\sin\pi\frac{l}{L}\right]^{x_\phi^s-x_\phi}
\label{offdiag}
\ee
which involves also the surface scaling dimension $x_\phi^s$.

\begin{figure}
\epsfxsize=\columnwidth\epsfbox{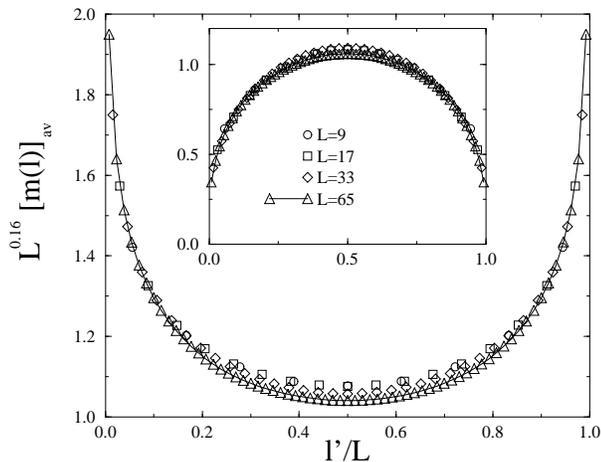}
\caption{Scaling plot of the magnetization profile $m_L(l)$ 
  with symmetric (fixed at both ends) boundary condition
  (\protect{\ref{diag}}) for the RS-sequence. The inset shows a
  scaling plot of the off-diagonal matrix element
  (\protect{\ref{offdiag}}) with symmetric (free at both ends)
  boundary condition.Here it $\lambda=4$, the bulk magnetization
  exponent is used as a fit parameter: $x_m=0.160(1)$ gives the best
  data collapse.}
\label{fig2}
\end{figure}

For the {\it random} TIM model several profiles have been calculated
in \cite{profiles,bigpaper} and they all follow very well the
conformal predictions. This coincidence of the numerical and the conformal
results is quite surprising, since the random TIM is not conformally
invariant due to strongly anisotropic scaling at the critical point.

Here we consider the same problem for aperiodic chains and calculate
the magnetization profiles for the RS-chain. In Fig.\ \ref{fig2} we
present the scaling plot of the diagonal and off-diagonal profiles
with symmetric boundary conditions. Here we take the exact value for
the surface magnetization scaling dimension $x_m^s$ from
eq(\ref{xmsrs}), whereas the bulk magnetization exponent, $x_m$, is
used as a fit parameter in order to obtain a good data collapse. As
seen in Fig.\ \ref{fig2} both profiles can be fitted with the same
exponent
\be
x_m(RS)=0.160(5)\;,
\ee
which turned out to be independent of the parameter $\lambda$ and
is different from the random chain value predicted by Fisher
\cite{fisher} to be $x_m({\rm rand})=(3-\sqrt{5})/4=0.191\ldots$. The
profiles, however, in analogy to the random case, follow very well the
conformal predictions, both for the diagonal and off-diagonal profiles
(\ref{diag}) and (\ref{offdiag}). This is an unexpected result, if we
take into account that the relevantly aperiodic Ising chains are not
conformally invariant, due to anisotropic scaling at the critical
point.

\subsection{Dynamical correlations}

Here we consider (imaginary) time dependent correlations of the same
spin, as defined in eq(\ref{autocorr}). One expects different types of
asymptotic behavior of the surface spins and of the bulk spins. First
we consider the bulk autocorrelation function
\be
G(\tau)=[\langle\sigma_{L/2}^x(\tau)\sigma_{L/2}^x\rangle]_{\rm av}
\label{corrdef}
\ee
and present a scaling consideration, where we essentially follow the
steps of reasoning in the random case \cite{riegerigloi,bigpaper}.

The autocorrelation function, like to the (local) magnetization, is
not self-averaging at the critical point: its average value is
determined by the {\it rare events}, which occur with a probability
$P_r$ and $P_r$ vanishes in the thermodynamic limit. In the random
quantum systems the disorder is strictly correlated along the time
axis, consequently in the rare events with a local order, i.e. with a
finite magnetization also the autocorrelations are
non-vanishing. Under a scaling transformation, when lengths are
rescaled as $l'=l/b$, with $b>1$ the probability of the rare events
transforms as $P_r'=b^{-x_m}$, like to the local magnetization. As we
said above the same is true for the autocorrelation function:
\be
G(\ln \tau)=b^{-x_m} G(\ln \tau /b^\omega)\qquad\delta=0\;,
\label{corrscale}
\ee
where we have made use of the relation between relevant time $t_r$ and
length $\xi$ at the critical point, which follows from the scaling
relation in eq(\ref{apdelta}). Setting the rescaling factor to 
$b=(\ln\tau)^{1/\omega}$ we obtain:
\be
G(\tau)\sim(\ln\tau)^{-x_m/\omega}\;.
\label{corrbulk}
\ee
For surface spins in (\ref{corrscale}) and (\ref{corrbulk}) one should
use the surface magnetization scaling dimension, $x_m^s$.

\begin{figure}
\epsfxsize=\columnwidth\epsfbox{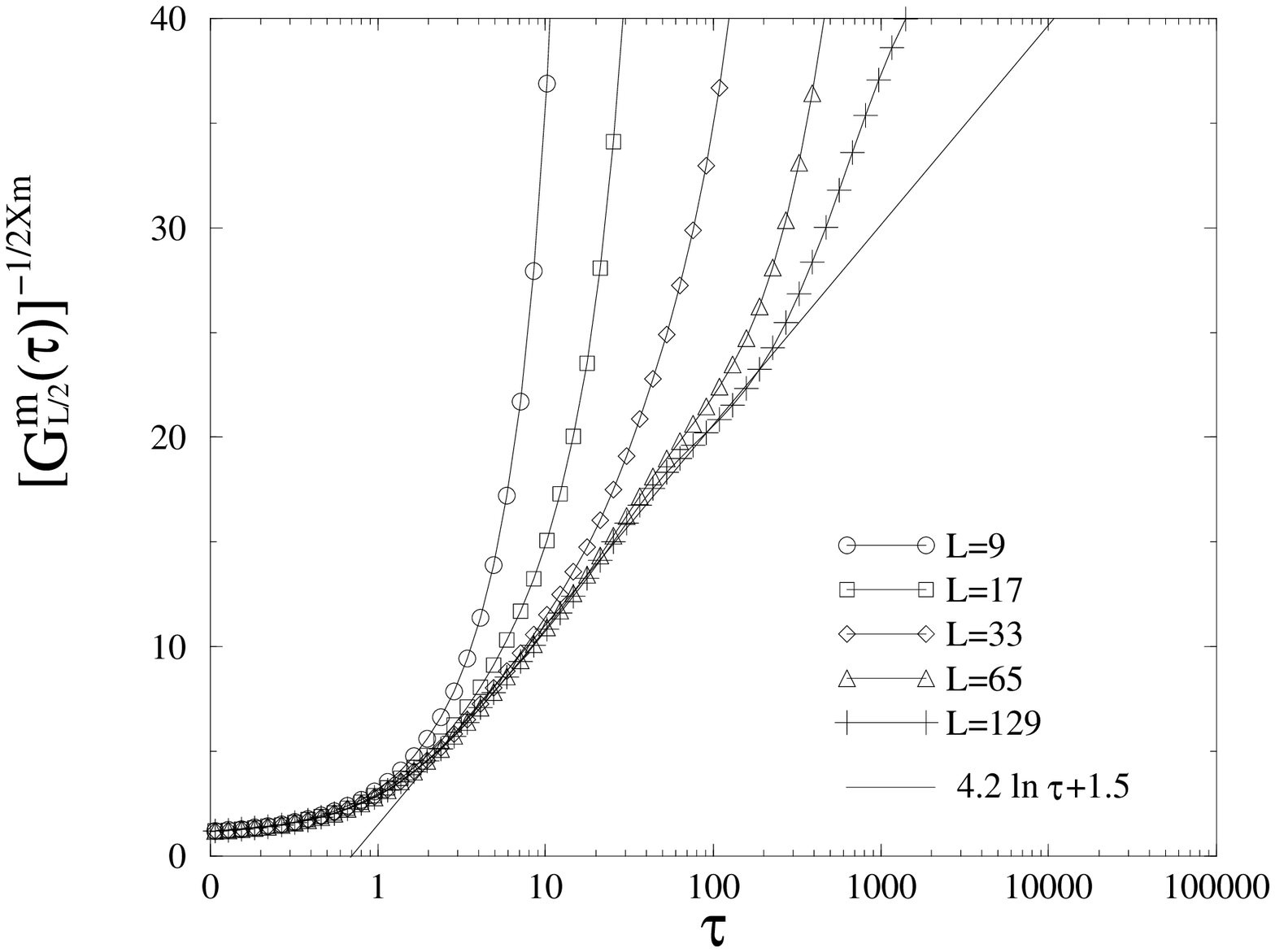}
\epsfxsize=\columnwidth\epsfbox{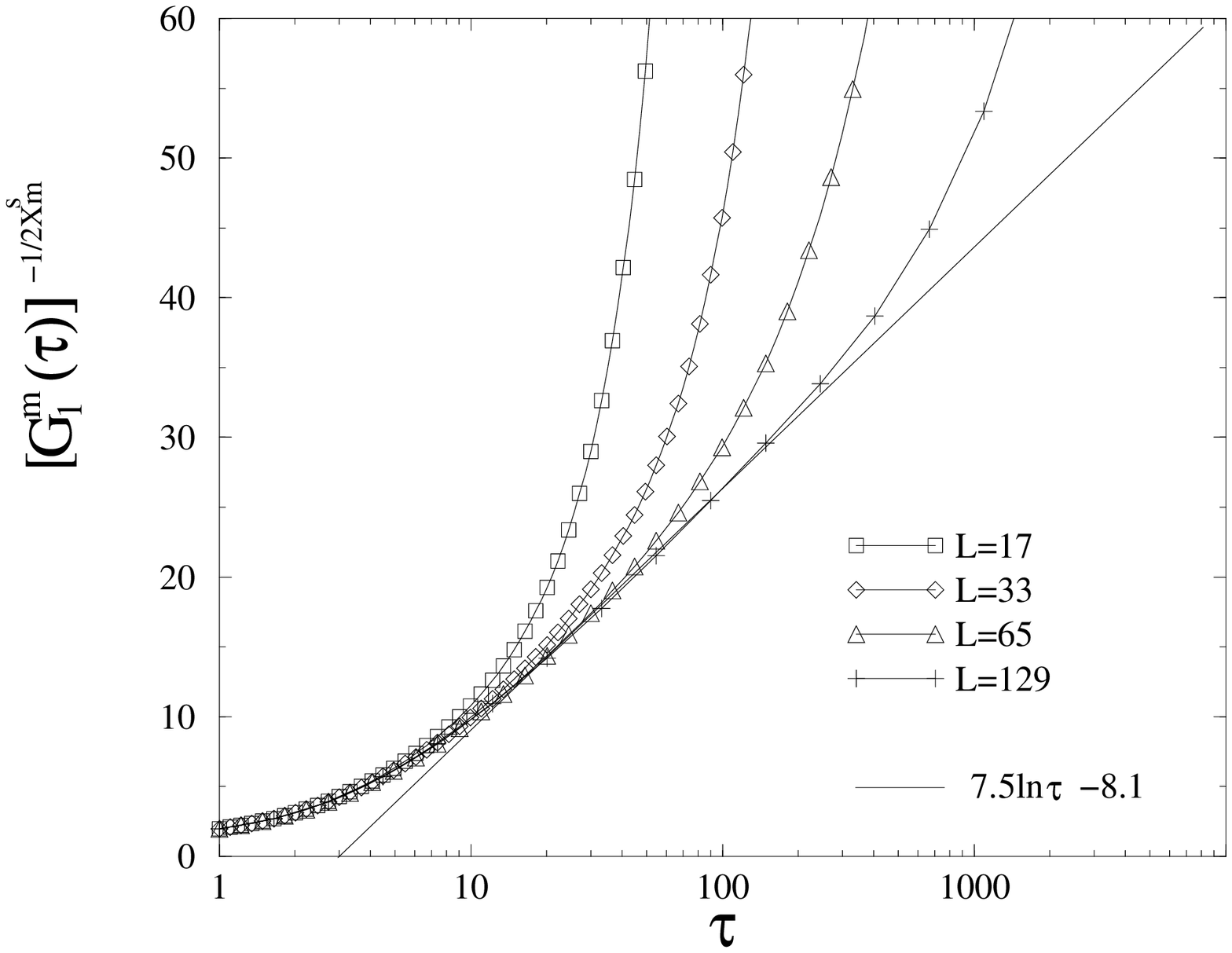}
\caption{
  {\bf a)} (Top): Bulk spin-spin autocorrelation function of the
  RS-sequence
  $G_{L/2}^m(\tau)=[\langle\sigma_{L/2}^x(t)\sigma_{L/2}^x\rangle]_{\rm
    av}$ in imaginary time for various system sizes (and $\lambda=4$).
  Note that we have chosen $L$ to be odd, so that $L/2$ denotes the
  central spin. In this plot with $[ G_{L/2}^m(\tau) ]^{-1/2x_m}$ on
  linear scale versus $\tau$ on a logarithmic scale the infinite
  system size limit is expected to lay on a straight line as
  indicated.  {\bf b)} (Bottom): Same as a) for the surface spin-spin
  autocorrelation function
  $G_1^m(\tau)=[\langle\sigma_1^x(\tau)\sigma_1^x\rangle]_{\rm av}$ in
  imaginary time.}
\label{fig3}
\end{figure}

We have numerically calculated the magnetization autocorrelation
function both at the surface and in the bulk for the RS-sequence. The
results are depicted in Fig.\ \ref{fig3}. As can be seen the finite
lattice results collapse onto one single curve (manifesting the
absence of finite size effects) and the critical temporal decay
happens on a logarithmic scale. The corresponding decay exponents are
given by $x_m/\omega$ ($x_m^s/\omega$), for bulk (surface)
correlations in agreement with the scaling result in (\ref{corrbulk}).

Next we consider the scaling behavior of the energy-energy
autocorrelation function
$G_l^e(\tau)=[\langle\sigma_l^z(\tau)\sigma_l^z\rangle]_{\rm av}$.  We
note that $\sigma_l^z$ represents one part of the local energy
operator, the other part of which --- $\sigma_l^x\sigma_{l+1}^z$ ---
is related to it through duality, see paper I. Therefore the above
autocorrelation function has essentially the same scaling behavior as
the full energy density.

\begin{figure}
\epsfxsize=\columnwidth\epsfbox{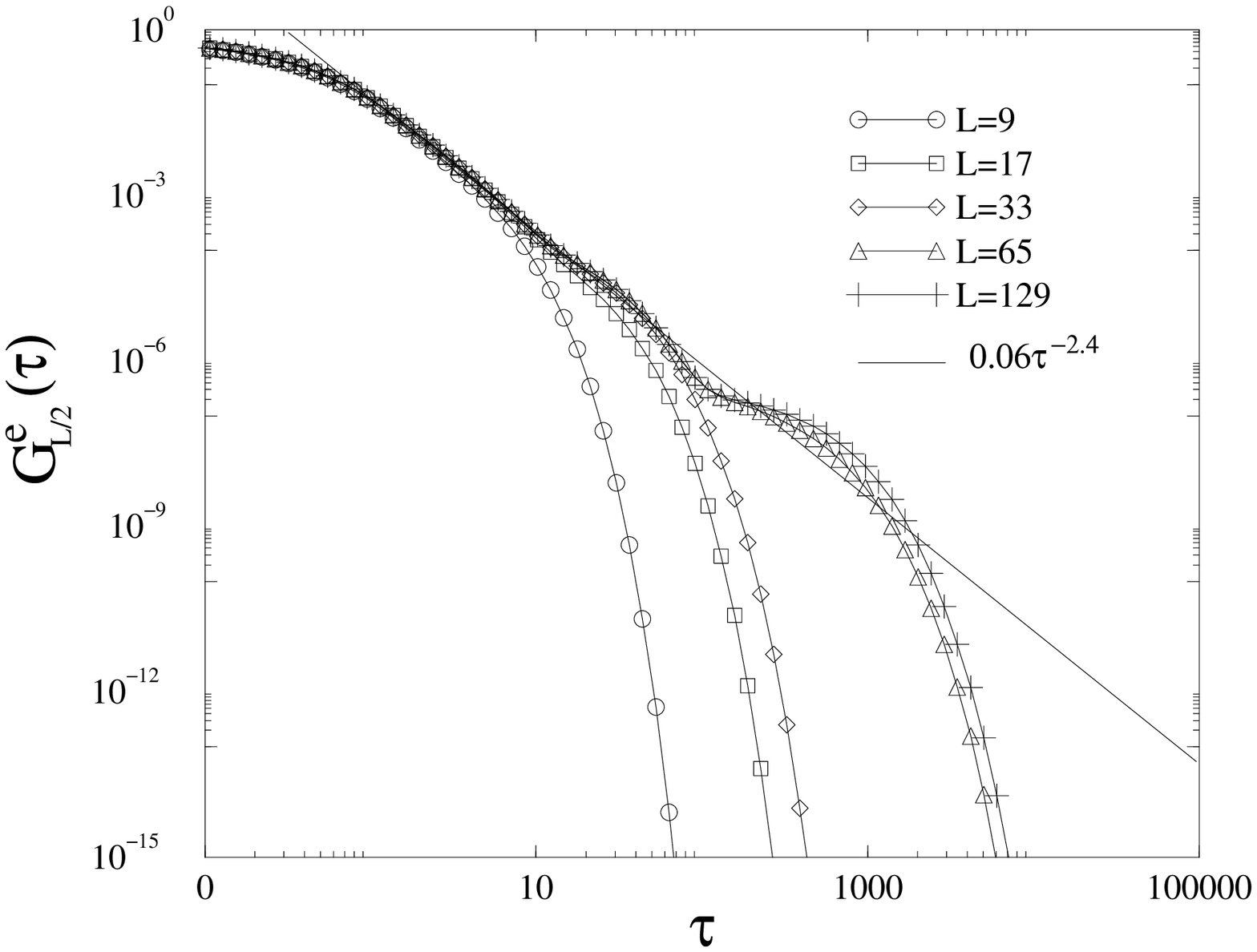}
\epsfxsize=\columnwidth\epsfbox{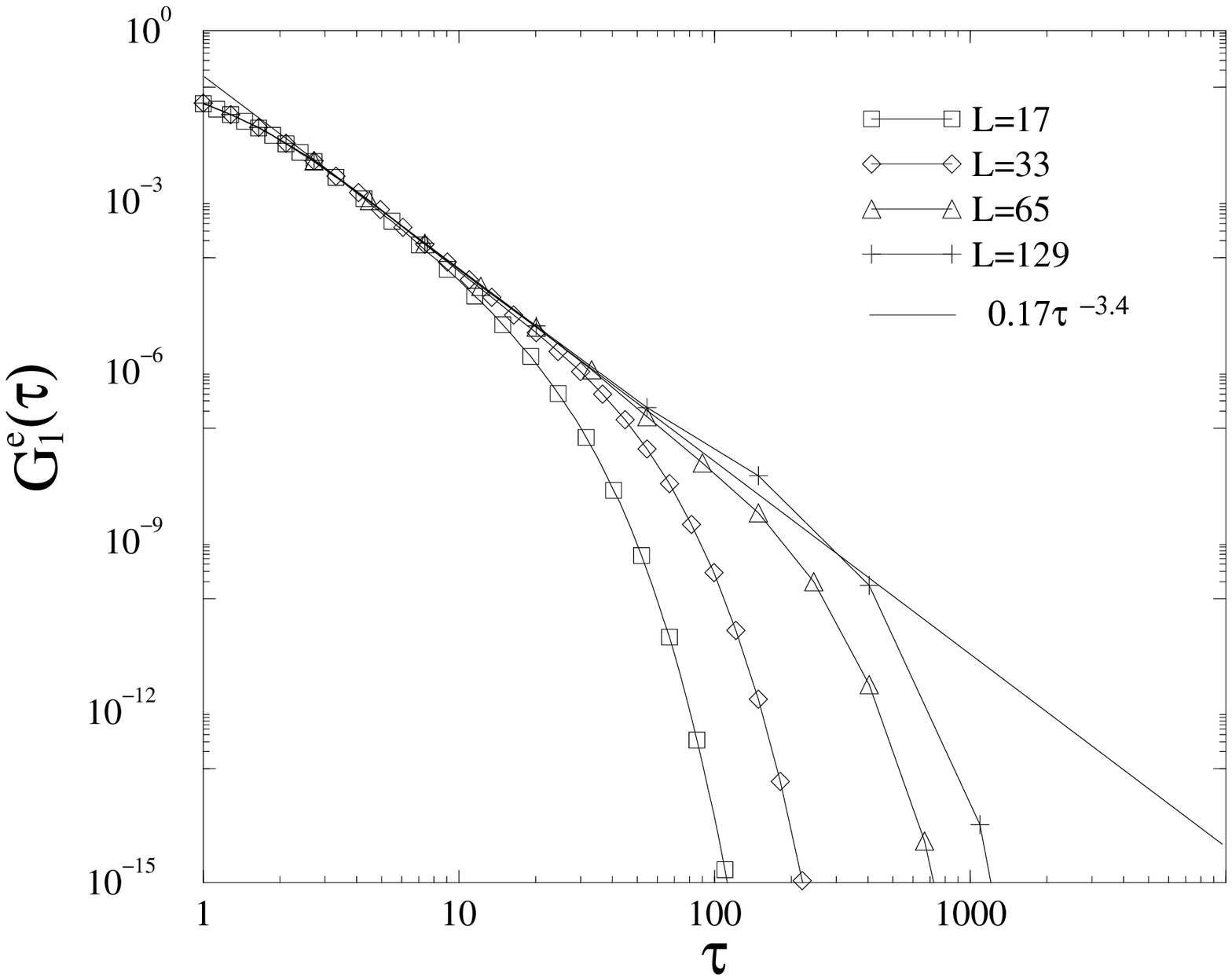}
\caption{
  {\bf a)} (Top): Bulk energy-energy autocorrelation function
  $G_{L/2}^e(\tau)=
  [\langle\sigma_{L/2}^z(\tau)\sigma_{L/2}^z\rangle]_{\rm av}$ in
  imaginary time for various system sizes (and $\lambda=4$) in a
  log-log plot. The straight line has slope $-2.4$, which yields our
  estimate for the exponent $\eta_e$.  {\bf b)} (Bottom): Same as a)
  for the surface energy-energy autocorrelation function
  $G_1^e(\tau)=[\langle\sigma_1^z(\tau)\sigma_1^z\rangle]_{\rm av}$ in
  imaginary time. The straight line has slope $-3.4$, which yields our
  estimate for the exponent $\eta_e^s$.}
\label{fig4}
\end{figure}

For the random chain, as was shown in I, the
critical energy autocorrelation function has an asymptotic power law
decay, 
\be
G_l^e(\tau)\sim\tau^{-\eta_e}
\ee
with the critical exponents $\eta_e({\rm rand})\approx2.2$ in the bulk
and $\eta_e^s({\rm rand})\approx2.5$ on the surface.  For the
aperiodic RS-chain the asymptotic decay is also consistent with a
power law decay, as can be seen in Fig.\ \ref{fig4}, both in the
bulk and at the surface of the system. The corresponding exponents
\be
\eta_e({\rm RS})\approx2.4
\qquad
\eta_e^s({\rm RS})\approx3.4
\ee
are, however, different from those of the random chain.

\section{Off-critical properties}

In random quantum systems there are Griffiths--McCoy singularities on
the paramagnetic side of the critical point, which result in a power
law decay of the autocorrelation function
$G(\tau)\sim\tau^{-1/z(\delta)}$, where the dynamical exponent
$z(\delta)$ is a continuous function of the control parameter
$\delta$. In the random TIM there is also a Griffiths-McCoy phase on
the ferromagnetic side of the critical point and the values of the
dynamical exponent in the two regions are related via duality, see
paper I.

\begin{figure}
\epsfxsize=\columnwidth\epsfbox{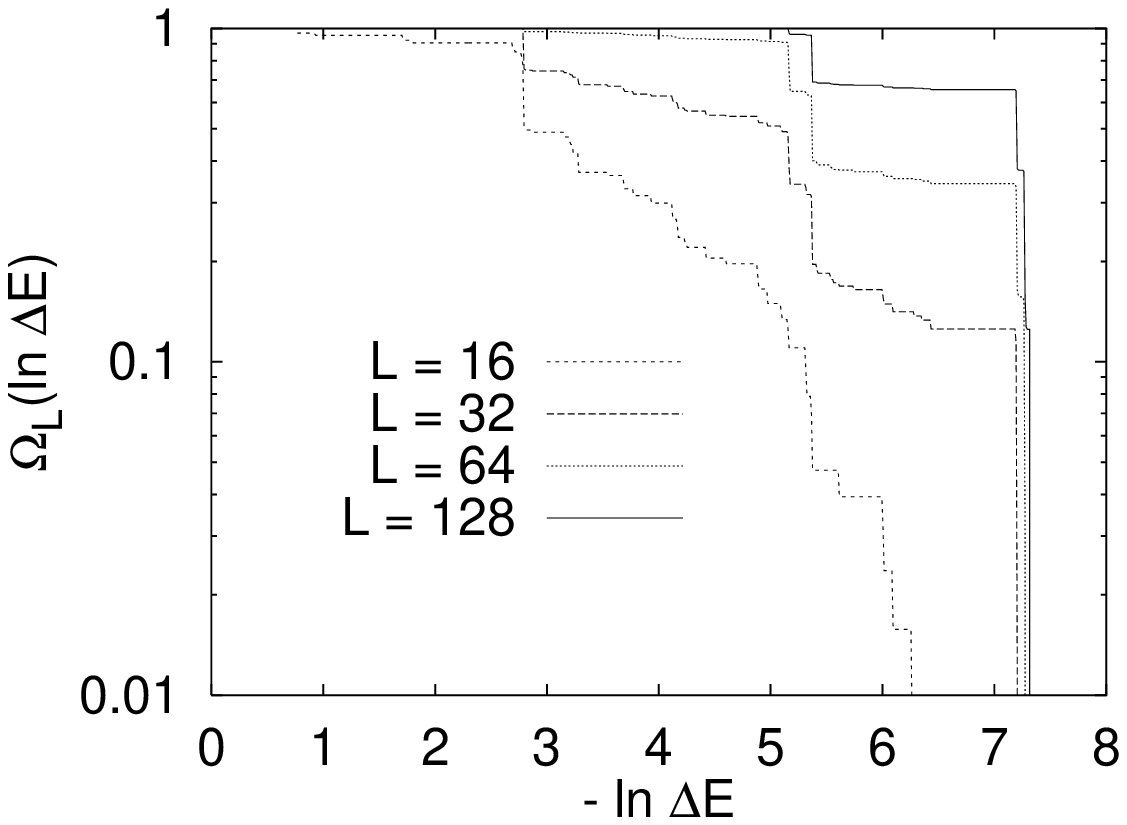}
\epsfxsize=\columnwidth\epsfbox{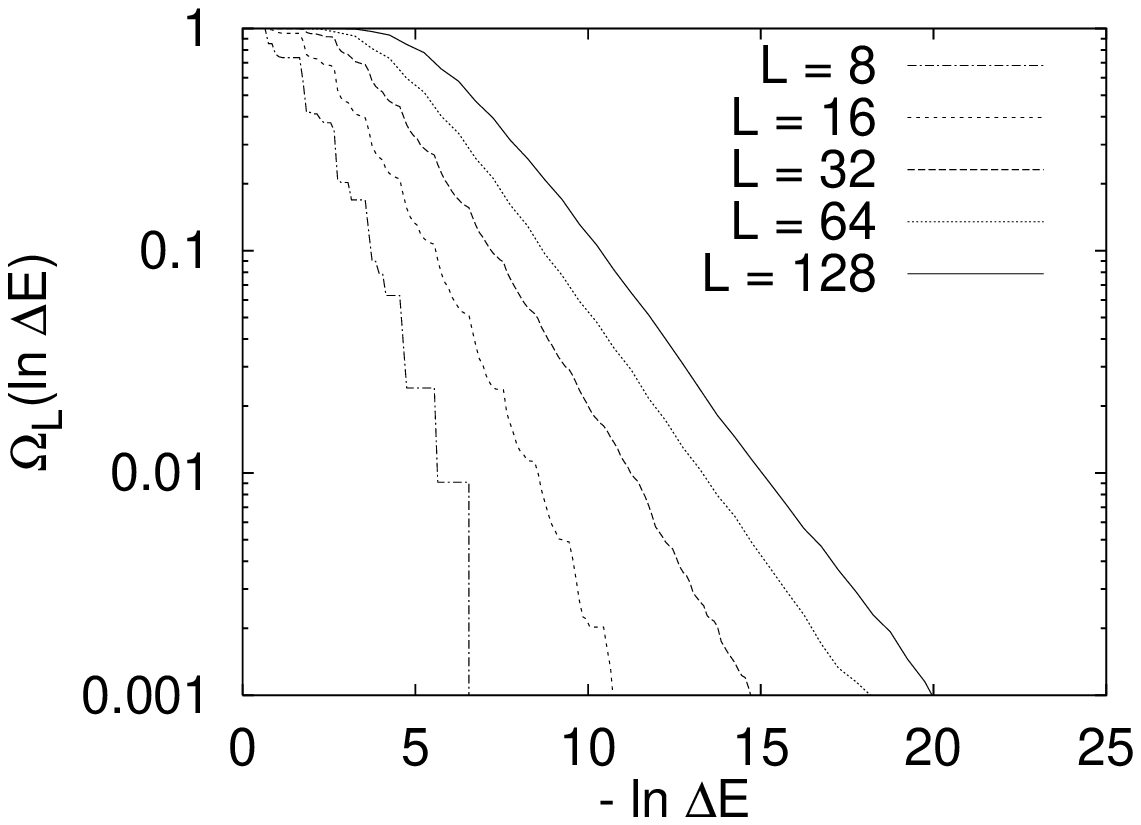}
\caption{{\bf a)} (Top) The integrated probability density 
  $\Omega(\ln\Delta E)$ for
  the RS sequence (exact average) with $\lambda=4$ slightly above the
  critical point ($h=1.5$). The distribution is chopped off at
  $\ln\Delta E_{\rm min}^{-1}(h=1.5)\approx7.3$. {\bf b)} (Bottom) The
  same as in a) for the {\it random} chain with binary disorder,
  $\lambda=4$. Note the larger $x$-range as compared with the
  aperiodic sequence and the absence of a cut-off for large enough
  system size. The asymptotic form of the distribution is 
  $\Omega(x)\sim\exp(x/z(\delta))$, where $x=\ln\Delta E$ and
  $1/z(h_0=1.5)=0.40$ (see paper I for details).}
\label{fig5}
\end{figure}

The long time behavior of the average autocorrelation function
$G(\tau)$ is determined by the Laplace transform of the gap
($\varepsilon_1$) distribution  function
\be
G(\tau)\sim\int_0^{\infty}d\varepsilon\,P(\varepsilon)
\exp(-\tau\varepsilon)
\ee
Thus the scaling properties of the low energy excitation are also
connected to the above defined dynamical exponent $z(\delta)$:
\be
\varepsilon(L,\delta)\propto L^{-z(\delta)}
\label{zgriff}
\ee
in a finite system of length $L$. 

For relevantly aperiodic chains the same type of scenario, i.e.\ the
existence of Griffiths-McCoy singularities, have been speculated
\cite{luck}. In the following we are going to clarify this issue and
study the distribution function of the energy gap $\varepsilon_1$ in
the disordered phase ($\delta>0$) of the RS-chain. In Fig.\ \ref{fig5}
we compare the integrated gap distribution functions for the random
and RS-chains. While the data for the random chain follow the scaling
prediction in (\ref{zgriff}) with $\delta=0.5\ln h\approx0.20$,
$z\approx2.5$, the probability distribution of the RS-chain is chopped off:
there is a $L$--independent cut-off at $\Delta E_{\rm min}(\delta)$.
Consequently there is a relevant time scale in the problem
$t_r\sim\Delta E_{\rm min}^{-1}(\delta)$ and the autocorrelation
function has an exponential decay. The susceptibility and other
physical quantities are analytic in the whole disordered phase, thus
there is {\it no Griffiths-McCoy region} in the RS-chain and we expect
a similar behavior for any other aperiodic quantum spin chain.

\begin{figure}
\epsfxsize=\columnwidth\epsfbox{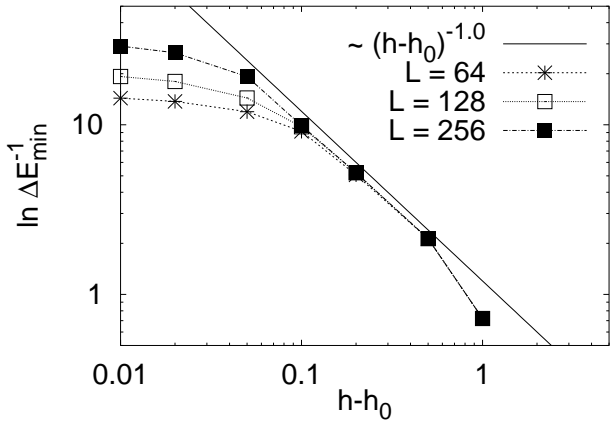}
\caption{
  The off-critical minimum energy gap of the RS-sequence
  $\ln\Delta E_{\rm min}^{-1}(h)$ versus the distance from the critical
  point ($h-h_0\sim\delta$ for $h\to h_0$) in the RS sequence for
  different system sizes. The straight line has slope $-1$, as
  predicted by (\protect{\ref{minigap}}).}
\label{fig6}
\end{figure}

We can estimate the minimum energy gap $\Delta E_{\rm min}(\delta)$ as
follows: We start with the formula for the excitation energy in
(\ref{epsform}) and consider a realization with surface order
$m_s={\cal O}(1)$ and $\overline{m}_s={\cal O}(1)$, which is generally
connected to the presence of a very small energy gap. Thus in the
paramagnetic phase $0<\delta\ll1$ we have 
\be
\Delta E_{\rm min}(\delta)\sim\varepsilon_1\sim
\prod_{i=1}^{L-1} {h_i\over J_i}
\sim\exp(A\delta L - BL^\omega)\;,
\label{deltaeq}
\ee
where $A>0$, $B>0$. The first term in the exponential describes the
average trend with $\delta$, whereas the second represents the largest
possible fluctuation in the couplings among all aperiodic sequences of
length $L$. It is important to note that for a random distribution
this second term could be proportional to $L$ in rare events, implying
that in the random case there is no minimum energy gap. For the
aperiodic chains, however, due to the number of different realizations
that increases only linearly with $L$ the fluctuating energy
(\ref{fluctuate}) grows slower than $L$. Consequently from
eq(\ref{deltaeq}) one can derive a length scale
\be
l_{\rm ap}\sim\delta^{-1/(1-\omega)}
\label{aplength}
\ee
that characterizes the most singular sample and the corresponding
minimum energy gap is then given by
\be
\Delta E_{\rm min}(\delta)\sim
\exp\left(-{\rm const}\cdot\delta^{-\omega/(1-\omega)}\right)\;.
\label{minigap}
\ee
This relation is indeed satisfied for the RS-chain, since according to
our numerical results in this case $\ln\Delta E_{\rm min}(\delta)\sim
1/\delta$, as can be seen in Fig.\ \ref{fig6}.

\section{Random walks in random and aperiodic environments}

As it was shown in paper I and being utilized in section IV/B in this
paper there is a close relation between the random quantum Ising spin
chains and the one-dimensional random walk. Especially the scaling
properties of the surface magnetization and that of the low energy
excitations of the random TIM can be obtained from the surviving
properties of a one-dimensional random walk. Here we go further and
emphasize a relation between the random TIM and the 1d random walk in
a random environment.

To be specific, we characterize the one-dimensional random walk with
the nearest neighbor hopping by the transition probabilities
$w_{i,j}=w(i\to j)$ for a random walker to jump from site $i$ to site
$j$ with
\be
w_{i,j}=\left\{
\begin{array}{ccc}
w_{i,i\pm1} & \quad{\rm for}\quad & \vert i-j\vert=1\\
      0     & \quad{\rm for}\quad & \vert i-j\vert>1
\end{array}
\right.\;.
\ee
Here we are particularly interested in the general case, in which the
transition probabilities are not necessarily symmetric, i.e.\ 
\be
w_{i,i+1}\ne w_{i+1,i}\;, 
\label{asym}
\ee
Moreover, the random walker is confined to a finite number of sites
$i=1,\ldots,L$. At the two ends of this interval, i.e.\ at $i=0$ and
$i=L+1$, we put {\it absorbing walls}, which is simply modeled by
setting $w_{0,1}=w_{L+1,L}=0$ (i.e.\ the walker cannot jump back into
the system once landed on $0$ or $L+1$).  The time evolution of the
probability distribution of the walk $P_{i,j}(t)$, which is the
probability for the walker to be at time $t$ on site $j$ once started
at time $0$ on site $i$, is fully determined by the Master-equation
\be
\frac{d}{dt}\,\,\underline{P}(t)=\underline{\underline{M}}\cdot
\underline{P}(t)\;.
\ee
Here
\be
\underline{P}(t)=
\left(P_{i,0}(t),P_{i,1}(t),\ldots,P_{i,L}(t),P_{i,L+1}(t)\right)^T
\ee
and the transition matrix is $(\underline{\underline{M}})_{i,j}=w_{i,j}$
for $i\ne j$ and $(\underline{\underline{M}})_{i,i}=-\sum_j w_{i,j}$
while the initial condition is $P_{i,j}(0)=\delta_{i,j}$. 
All physical properties of the model can be expressed in terms of
(left and right) eigenvectors and eigenvalues of
$\underline{\underline{M}}$, very much in the same way as the physics
of the TIM is contained in the eigenvectors and eigenvalues of the
tridiagonal matrix (\ref{trid}).

Here we consider first one quantity that gained considerable interest
recently in related models for anomalous diffusion
\cite{derrida,persistence}: The {\it persistence probability} $P_{\rm
  pr}(L,t)$, which is the probability that a walker starting at site
$i=1$ does not cross its starting point until time $t$. Due to the
absorbing sites at $i=0$ and $i=L+1$ its long time limit $p_{\rm
  pr}(L)=\lim_{t\to\infty}P_{\rm pr}(L,t)$ is simply given by
\cite{walks}
\be
p_{\rm pr}(L)
=\lim_{t\to\infty}P_{1,L+1}(t)
=\left(1+\sum_{i=1}^L\prod_{j=1}^i
\frac{w_{j,j-1}}{w_{j,j+1}}\right)^{-1}
\label{persl}
\ee
Thus, as is shown in \cite{walks}, there is a one-to-one relation
between the persistence probability (\ref{persl}) and the surface
magnetization $m_s(L)$ of the TIM (\ref{surfmag}) with the following
correspondences
\be
\begin{array}{rcl}
w_{i,i+1} & \longrightarrow & J_i^2\\
w_{i,i-1} & \longrightarrow & h_i^2\\
p_{\rm pr}(L) & \longrightarrow & m_s^2(L)
\end{array}
\label{corresp}
\ee
Consequently similar relations hold for the average quantities, when
the transition probabilities (or equivalently the fields and the
couplings) follow the same random or aperiodic modulation.

In the random case the critical point of the TIM corresponds to the
Sinai walk \cite{sinai}, and from eq(\ref{surfmag}), (\ref{xmsrand})
and (\ref{persl}) we have
\be
[p_{\rm pr}^{\rm (rand)}(L)]_{\rm av}
\propto L^{-1/2}\;.
\ee
For relevantly aperiodic environments which ar characterized by a
wandering exponent $\omega>0$ we get from (\ref{xmsall})
\be
[p_{\rm pr}^{\rm (aperiodic)}(L)]_{\rm av}
\propto L^{-(1-\omega)}\;.
\ee
In the non-critical situation there is an average drift of the walk,
which can be defined through eq(\ref{delta}) as 
\be
\delta_{\rm RW}={[\ln w_\rightarrow]_{\rm av}-[\ln w_\leftarrow]_{\rm av} 
\over \rm{var}[\ln w_\rightarrow]+\rm{var}[\ln w_\leftarrow]}\;,
\label{deltarw}
\ee
where $w_\rightarrow$ ($w_\leftarrow$) stands for transition
probabilities to the right (left), i.e.\ $w_{i,i+1}$ ($w_{i,i-1}$).
For $\delta_{\rm RW}\ne 0$ the average correlations defined on persistent
walks are characterized by a correlation length
\be
\xi\sim\vert\delta_{\rm RW}\vert^{-\nu}\;,
\ee
with the exponents for the {\it average} given in eq(\ref{nurand}) and
eq(\ref{exprel}) for the random and aperiodic environments,
respectively.

The dynamical properties of the random walk are dominated by the
largest, non-vanishing eigenvalue $\lambda_{\rm m}$ of the
transition matrix $\underline{\underline{M}}$. It can be shown
\cite{walks} that for $\lambda_{\rm m}$ a formula similar to
eq(\ref{epsform}) holds, where $h_i$ and $J_i$ have to be replaced by
their random walk counterparts given in (\ref{corresp}) and for the
term $m_s \overline{m_s}$ stands the persistence probability
$p_{\rm pr}(L)$. This implies
straightforwardly, using eq(\ref{apdelta}) that in a Sinai walk in the
random and relevantly aperiodic case the characteristic time scales
like
\be
t_{\rm char}^{\rm(rand)}
\sim\Bigl(\lambda_{\rm m}^{\rm(rand)}\Bigr)^{-1}
\sim\exp({\rm const}\cdot L^{\omega})
\ee
with $\omega=1/2$ in the random case. As a further consequence,
autocorrelations or return probabilities decay logarithmically, 
\be
P_0(t)=\frac{1}{L}\sum_{i=1}^L P_{i,i}(t)
\propto\frac{1}{\ln^\eta(t)}
\ee
with $\eta=2$ in the random case and $\eta=1/\omega$ in the relevantly
aperiodic case. This follows from the scaling result $P_0(t)^2 \sim
\left[X^2(t)\right]_{\rm av}^{-1}$, where $\left[X^2(t)\right]_{\rm av}$
is the average mean-square displacement, as already mentioned in the
Introduction.

Finally the Griffiths-McCoy phase of the {\it random} TIM is
equivalent to the anomalous diffusion region of the random walk with
$\delta_{\rm~RW}\ne0$, in which case autocorrelations decay
anomalously slow with an exponent $\gamma(\delta_{\rm RW})\le1$ 
\be
P_0(t)\sim t^{-\gamma(\delta_{\rm RW})}
\ee
that depends continuously on the drift parameter $\delta_{\rm RW}$ and
corresponds to the inverse dynamical exponent $z(\delta)$ of the
Griffiths-McCoy phase of the random TIM. In the two limiting cases we
have $z(\delta)=1/2 \delta$ as $|\delta| \to 0$\cite{fisher} and $z(\delta)=1$
as $|\delta| \to \infty$. In the latter case we approach the ballistic
situation, when all the steps of the walk are made in the same direction, thus
the average displacement has a linear time dependence.
We can thus extend our dictionary (\ref{corresp}) by
\be
\begin{array}{rcl}
\lambda_{\rm m}       & \longrightarrow & \varepsilon_1\\
\gamma(\delta_{\rm RW}) & \longrightarrow & 1/z(\delta)\\
\end{array}
\label{corresp2}
\ee
For the {\it aperiodic} case with drift we remember the results of
section V, in particular eq(\ref{minigap}) and conclude that a random
walk in a relevantly aperiodic environment for $\delta_{\rm RW}\ne0$
does {\it not} exhibit a region of anomalous diffusion.

\section{Summary}

To summarize we have studied the effect of random and aperiodic
environments on cooperative processes in one space dimension. We have
shown that at the critical point, both for the transverse-field Ising
model and for the diffusion process, the two types of inhomogeneities
have quite similar consequences, which is based on the same type of
distribution of the low energy excitations (large time scales). We
have obtained - presumably exact - scaling relations, which connect
the values of the surface magnetization exponent and that of the
correlation length exponent with the known characteristics of the
(random and aperiodic) environments. Besides the similarities between
the critical properties of random and aperiodic models we have also
observed several quantitative differences. For example some critical
exponents are turned out to be environment dependent and --- most
noticeably --- the Griffiths phase is absent for aperiodic models.

\acknowledgements

Two of us F.\ I. and H.\ R. would like to express their deep gratitude
to Prof. J. Zittartz for many stimulating discussions and for his
efforts to provide the right conditions for their research work, of
which, among others, the present collaboration is a result.

This work has been supported by the French-Hungarian cooperation
program "Balaton" (Minist\`ere des Affaires Etrang\`eres-O.M.F.B), the
Hungarian National Research Fund under grants No OTKA TO12830, OTKA
TO23642 and OTKA TO25139 and by the Ministery of Education under grant
No FKFP 0765/1997. F.\ I. thanks the HLRZ in KFA J\"ulich, where part
of this work has been completed, for kind hospitality.  H.\ R.'s work
was supported by the Deutsche Forschungsgemeinschaft (DFG). The
Laboratoire de Physique des Materiaux is Unit\'e Mixte de Recherche
CNRS No 7556.

\appendix
\section*{Average critical surface magnetization for the Rudin-Shapiro chain}

We consider the RS-chain in (\ref{rs}) generated from a letter {\bf A}
and calculate the surface magnetization (\ref{surfmag}) for chains of
length $L=2^{2l+1}$, $l=1,2,\ldots$. The chain starts at the $1$st,
$2$nd,$\ldots$,$N$th position of the original RS-chain, such that
$N=L\cdot2^{2k}$, $k=1,2,\ldots$, and average over these $N$
realizations. The average critical surface magnetization in the limit
$\lambda\to\infty$ is given by:
\end{multicols}
\renewcommand{\theequation}{\Alph{section}\arabic{equation}}
\widetext
\noindent
%
\be
m_s(l,k)=\frac{1}{N}\biggl\{
N_1(l,k)
+\frac{N_2(l,k)}{\sqrt{2}}
+\frac{N_u(l,k)}{\sqrt{2^k+1}}
+H(k)[S_1(l)+S_2(l)]\biggr\}
\ee
where
\beqn
S_1(l)&=& \sum_{m=1}^{l-1} 2(l-m)
          \sum_{n=1}^{2^{m-1}}\frac{1}{\sqrt{2^{m-1}+1+n}}\\
S_2(l)&=& \sum_{n=1}^{2^{l-1}-1}\frac{1}{\sqrt{2^{k-1}+1+n}}\\
N_1(l,k)&=& 18\cdot4^{k-2}+9\cdot2^{k-2}+5(2^{l-1}-1)(4^{k-1}+2^{k-1})\\
N_2(l,k)&=& 4^{k-2}\Bigl(10\cdot2^{l-1}+4(l-1)\Bigr)
           +\Bigl(5\cdot2^{l-1}+2(l-1)-2\Bigr)\cdot2^{k-2}\\
N_u(l,k)&=& 2^{2k-1}+(4^{k-1}-2^{k-1})(2^{l-1}-1)\\
H(k)&=& 2\cdot4^{k-2}+2^{k-2}
\eeqn
The asymptotic behavior of $\lim_{k\to\infty} m_s(l,k)$ is given in
eq(\ref{rslimit}).
\begin{multicols}{2} 
\narrowtext

\end{multicols}


\noindent


\begin{references}



\bibitem{sinai}
        Ya.\ G.\ Sinai, Theor. Probab. Appl. {\bf 27}, 247 (1982);
        For a review on random walks in disordered media see:
        J.\ P.\ Bouchaud and A.\ Georges, 
        Phys.\ Rep.\ {\bf 195}, 127 (1990)

\bibitem{bigpaper}
        F. Igl\'oi and H.\ Rieger, Phys. Rev. B (in press).

\bibitem{riegerigloi}
        H. Rieger and F. Igl\'oi,
        Europhys. Lett. {\bf 39}, 135 (1997).

\bibitem{griffiths}
        R.B. Griffiths, Phys. Rev. Lett. {\bf 23}, 17 (1969).

\bibitem{mccoy}
        B. McCoy, Phys. Rev. Lett. {\bf23}, 383 (1969).

\bibitem{aperiodicshort}
        F. Igl\'oi, D. Karevski and H. Rieger,
        Eur.\ Phys.\ J.\ B {\bf 1}, (1998).        

\bibitem{fisher}
        D.S. Fisher, Phys. Rev. Lett. {\bf 69}, 534 (1992); 
        Phys. Rev. B {\bf 51}, 6411 (1995).

\bibitem{mckenzie}
        R. H. McKenzie, 
        Phys. Rev. Lett. {\bf 77}, 4804 (1996).
         
\bibitem{youngrieger}
        A. P. Young and H. Rieger, 
        Phys. Rev. B {\bf 53}, 8486 (1996).

\bibitem{profiles}
        F. Igl\'oi and H.\ Rieger,
        Phys. Rev. Lett. {\bf 78}, 2473 (1997).

\bibitem{young}
        A. P. Young, Phys. Rev. B {\bf 56}, 11691 (1997).

\bibitem{qsg} 
        See H. Rieger and A. P Young, in {\it Complex Behavior
        of Glassy Systems}, ed.\ M. Rubi and C. Perez-Vicente, Lecture
        Notes in Physics {\bf 492}, p.\ 256, Springer-Verlag,
        Heidelberg, 1997, for a review on the Ising quantum spin glass
        in a transverse field.

\bibitem{senthil}
        S. Sachdev and T. Senthil,
        Phys. Rev. Lett. {\bf 77}, 5292 (1996).

\bibitem{fm2d}
        H.\ Rieger and N.\ Kawashima, submitted to Phys. Rev. Lett.; 
        C.\ Pich and A.\ P.\ Young, submitted to Phys. Rev. Lett.;
        T.\ Ikegami, S.\ Miyashita and H.\ Rieger, J. Phys. Soc. Jap.
        (in press).

\bibitem{surfising}
        B. M. McCoy and T. T. Wu,
        Phys. Rev. {\bf 162}, 436 (1967).

\bibitem{mccoywu}
        B.M. McCoy and T.T. Wu, 
        Phys. Rev. {\bf 176}, 631 (1968); {\bf 188}, 982 (1969);
        B.M. McCoy, Phys. Rev. {\bf 188}, 1014 (1969).

\bibitem{zittartz}
        P. Hoever, W. F. Wolff and J. Zittartz, 
        Z. Phys. B {\bf 41}, 43 (1981);
        W. F. Wolff, P. Hoever and J. Zittartz, 
        Z. Phys. B {\bf 42}, 259 (1981);
        P. Hoever and J. Zittartz, 
        Z. Phys. B {\bf 44}, 129 (1981).

\bibitem{shankar}
        R. Shankar and G. Murthy, 
        Phys. Rev. B {\bf 36}, 536 (1987).

\bibitem{nieuwenhuizen}
        T. M. Nieuwenhuizen and H. Orland,
        Phys. Rev. B {\bf 40}, 5094 (1989).

\bibitem{harris}
        A.B. Harris, J. Phys. C {\bf 7}, 1671 (1974).

\bibitem{luck}
        J.M. Luck, J. Stat. Phys. {\bf 72}, 417 (1993).

\bibitem{igloi93}
        F. Igl\'oi, J. Phys. A {\bf 26}, L703 (1993).

\bibitem{dekking}
        M. Dekking, M. Mend\'es-France and A. van der Poorten,
        Math. Intelligencer, {\bf 4}, 130 (1983)               

\bibitem{bercheberche}
        B. Berche et al: J. Phys. A {\bf 28}, L165 (1995).

\bibitem{igloiturban96} 
        F. Igl\'oi and L. Turban, 
        Phys. Rev. Lett. {\bf 77}, 1206 (1996).

\bibitem{itks}
        F. Igl\'oi, L. Turban, D. Karevski and F. Szalma, Phys.
        Rev. B {\bf 56}, 11031 (1997)

\bibitem{grimmbaake}
        J. Hermisson, U. Grimm and M. Baake
        J. Phys. A {\bf 30}, 7315 (1997).

\bibitem{igloiturban94}
        F. Igl\'oi and L. Turban, Europhys. Lett. {\bf 27}, 91 (1994).

\bibitem{bcb}
        P.-E. Berche, C. Chatelain, B. Berche, Phys. Rev. Lett. {\bf
        80}, 297 (1998)

\bibitem{fermion}
        E. Lieb, T. Schultz and D. Mattis, 
        Ann. Phys. (N.Y.) {\bf 16}, 407 (1961);
        P. Pfeuty, Ann. Phys. (Paris) {\bf 57}, 79 (1970).

\bibitem{BST}
        R. Bulirsch and J. Stoer, Num. Math. {\bf 6}, 413 (1964) 

\bibitem{fisherdegennes}
        M.E. Fisher and P.-G. De Gennes, 
        C.R. Acad. Sc. Paris B {\bf 287}, 207 (1978).

\bibitem{burkhardt}
        T.W. Burkhardt and T. Xue, 
        Phys. Rev. Lett. {\bf 66}, 895 (1991);
        Nucl. Phys. {\bf B354}, 653 (1991)

\bibitem{turbanigloi97}
        L. Turban and F. Igl\'oi, J. Phys. A {\bf 30}, L105 (1997).

\bibitem{derrida}
        B. Derrida, A.J. Bray and C. Godr\`eche, 
        J. Phys. A {\bf 27}, L357 (1994);
        B. Derrida, V. Hakim and V. Pasquier, 
        Phys. Rev. Lett. {\bf 75}, 751 (1995).

\bibitem{persistence}
        S.\ N.\ Majumdar and C.\ Sire,
        Phys.\ Rev.\ Lett.\ {\bf 77}, 1420 (1996);
        S.\ N.\ Majumdar, C.\ Sire, A. J. Bray and S. J. Cornell,
        Phys.\ Rev.\ Lett.\ {\bf 77}, 2867 (1996);
        S.\ N.\ Majumdar, A. J. Bray, S. J. Cornell, and C.\ Sire,
        Phys.\ Rev.\ Lett.\ {\bf 77}, 3704 (1996);
        K.\ Oerding, S.\ J.\ Cornell, and A.\ Bray,
        Phys.\ Rev.\ E {\bf 56}, R25 (1997);
        J. Krug, H. Kallabis, S.\ N.\ Majumdar, S.\ J.\ Cornell,
        A. J. Bray,  and C.\ Sire,
        Phys. Rev. E {\bf 56}, 2702 (1997).


\bibitem{walks}
        F. Igl\'oi and H. Rieger, to be published.

\end{references}
\end{document}